\documentclass[aps,prd,reprint,superscriptaddress,notitlepage,%nofootinbib,
nobibnotes,amsmath,amssymb,preprintnumbers]{revtex4-2}

\usepackage{graphicx}% Include figure files
\usepackage{dcolumn}% Align table columns on decimal point
\usepackage{bm}% bold math
\usepackage{hyperref}% add hypertext capabilities
\usepackage{comment}

\newcommand{\refeq}[1]{Eq.~(\ref{eq:#1})}

\newcommand{\be}{\begin{equation}}
\newcommand{\ee}{\end{equation}}
\newcommand{\bea}{\begin{eqnarray}}
\newcommand{\eea}{\end{eqnarray}}
 
\def\ba#1\ea{\begin{align}#1\end{align}}

\usepackage{xcolor}
\definecolor{RedWine}{rgb}{0.743,0,0}

\begin{document}

\preprint{APCTP-Pre2025-005}

\title{Planck isocurvature constraint on primordial black holes lighter than a kiloton}% Force line breaks with \\

\author{TaeHun Kim}
\email{gimthcha@kias.re.kr}
\affiliation{School of Physics, Korea Institute for Advanced Study, Seoul 02455, Korea}

\author{Jinn-Ouk Gong}
%\email{jgong@ewha.ac.kr}
\affiliation{Department of Science Education, Ewha Womans University, Seoul 03760, Korea}
\affiliation{Asia Pacific Center for Theoretical Physics, Pohang 37673, Korea}

\author{Donghui Jeong}
%\email{djeong@psu.edu}
\affiliation{Department of Astronomy and Astrophysics and Institute for Gravitation and the Cosmos, 
The Pennsylvania State University, University Park, Pennsylvania 16802, USA}
\affiliation{School of Physics, Korea Institute for Advanced Study, Seoul 02455, Korea}

\author{Dong-Won Jung}
%\email{dongwon.jung@ibs.re.kr}
\affiliation{Cosmology, Gravity and Astroparticle Physics Group,
Center for Theoretical Physics of the Universe,
Institute for Basic Science, Daejeon, 34126, Korea}

\author{Yeong Gyun Kim}
%\email{ygkim@gnue.ac.kr}
\affiliation{Department of Science Education, Gwangju National University of Education, Gwangju 61204, Korea}

\author{Kang Young Lee}
%\email{kylee.phys@gnu.ac.kr}
\affiliation{Department of Physics Education \& Research Institute of Natural Science, Gyeongsang National University, Jinju 52828, Korea}


\begin{abstract}
We demonstrate that primordial black holes (PBHs) lighter than $10^9 \, \text{g}$, which evaporated before the big bang nucleosynthesis, can induce significant isocurvature perturbations due to their biased clustering amplitude and the branching ratio of the Hawking radiation differing from the abundance ratio. By leveraging the upper bound on the isocurvature perturbations from the cosmic microwave background anisotropies reported by the Planck collaboration, we derive a new upper bound on the abundance of these light PBHs in the presence of primordial non-Gaussianity as a working example.

\end{abstract}

\maketitle

%%%%%%%%%%%%%%%%%%%%%%%%%%%%%%%%%%%%%%%%%%%%%%%%%%%%%%%%%%%%%%%%%%%%%%%%%%%%%%%
%%%%%%%%%%%%%%%%%%%%%%%%%%%%%%%%%%%%%%%%%%%%%%%%%%%%%%%%%%%%%%%%%%%%%%%%%%%%%%%
%%%%%%%%%%%%%%%%%%%%%%%%%%%%%%%%%%%%%%%%%%%%%%%%%%%%%%%%%%%%%%%%%%%%%%%%%%%%%%%
\section{Introduction}
%%%%%%%%%%%%%%%%%%%%%%%%%%%%%%%%%%%%%%%%%%%%%%%%%%%%%%%%%%%%%%%%%%%%%%%%%%%%%%%
%%%%%%%%%%%%%%%%%%%%%%%%%%%%%%%%%%%%%%%%%%%%%%%%%%%%%%%%%%%%%%%%%%%%%%%%%%%%%%%
%%%%%%%%%%%%%%%%%%%%%%%%%%%%%%%%%%%%%%%%%%%%%%%%%%%%%%%%%%%%%%%%%%%%%%%%%%%%%%%

Primordial black holes (PBHs)~\cite{Zeldovich:1967lct, Carr:1974nx, Carr:1975qj} are hypothetical spacetime singularities that could have been produced in the early Universe through large primordial perturbations, generated by inflation~\cite{Yokoyama:1995ex, Garcia-Bellido:1996mdl, Garcia-Bellido:2017mdw}, first-order phase transitions~\cite{Hawking:1982ga, Moss:1994iq, Khlopov:1998nm, Jung:2021mku, Hong:2020est, Kawana:2021tde, Lu:2022paj, Kawana:2022lba, Lu:2022jnp, Lu:2022yuc, Marfatia:2024cac, Liu:2021svg, Kawana:2022olo}, and other scenarios~\cite{Hawking:1987bn, Cotner:2018vug, Conzinu:2020cke, Ruffini:1969qy, Cotner:2017tir, Cotner:2019ykd, Amendola:2017xhl, Flores:2020drq}. Although first predicted several decades ago, LIGO's discovery of gravitational waves in 2015 reignited interest in the cosmological and astrophysical consequences of PBHs~\cite{Bird:2016dcv}.

Once formed, PBHs lose mass and eventually evaporate through Hawking radiation, emitting all particles participating in gravity, including photons, baryons, as well as dark matter (DM) particles~\cite{Hawking:1974rv, Hawking:1975vcx, Hawking:1976de, Page:1976df, Page:1976ki, Page:1977um}. PBHs with masses $M_\text{PBH} \gtrsim 10^{15} \, \text{g}$ survive until the present time and can therefore play important cosmological roles. For example, they can occupy a significant fraction of cold DM for $10^{17} \, {\rm g} \lesssim M_\text{PBH} \lesssim 10^{22} \, {\rm g}$~\cite{Carr:2020xqk}. Outside this mass window, extensive searches on PBHs constrain the upper limit of their abundances~\cite{Carr:2020gox}. For $M_{\rm PBH}\gtrsim10^{22} \, \text{g}$, the gravitational influences of PBHs give constraints from gravitational microlensing ~\cite{Macho:2000nvd, Wilkinson:2001vv, EROS-2:2006ryy, Griest:2013esa, Griest:2013aaa, Niikura:2017zjd, Oguri:2017ock, Zumalacarregui:2017qqd, Niikura:2019kqi, Smyth:2019whb}, gravitational wave~\cite{Saito:2008jc, Ali-Haimoud:2017rtz, LIGOScientific:2019kan, Vaskonen:2019jpv, Chen:2019xse}, gas accretion~\cite{Inoue:2017csr, Serpico:2020ehh, Lu:2020bmd}, and astrophysical dynamics~\cite{Monroy-Rodriguez:2014ula, Koushiappas:2017chw, Gong:2017sie, Carr:2018rid, MUSE:2020qbo}. For $M_\text{PBH}\lesssim10^{17} \, {\rm g}$, constraints from the Hawking radiation come in. PBHs lighter than $10^{15} \, {\rm g}$ are considered to have completely evaporated by now, so their past abundances are constrained by the cosmic microwave background (CMB)~\cite{Carr:2009jm, Acharya:2020jbv, Chluba:2020oip}, $\gamma$-rays~\cite{Carr:2009jm, Carr:2016hva, DeRocco:2019fjq, Laha:2019ssq, Laha:2020ivk, Coogan:2020tuf}, cosmic rays~\cite{Boudaud:2018hqb, Dasgupta:2019cae}, radio waves~\cite{Chan:2020zry}, gas heating~\cite{Kim:2020ngi}, and light element abundances~\cite{Carr:2009jm}. For PBHs lighter than $10^9 \, \text{g}$, however, their lifetime is very short, and evaporation is completed before big bang nucleosynthesis (BBN), and no direct observational constraints have been known so far.

In this paper, we place new constraints on the abundance of these light PBHs that have evaporated well before BBN. Our constraints rely on two observations. First, the spatial distribution of PBHs can differ from that of other energy components; that is, PBHs can be ``biased'' tracers. This is because PBHs form only in rare regions with exceptionally large primordial perturbations. Second, the composition of Hawking radiation follows the effective Hawking degrees of freedom and generally differs from the relic abundance. Because of these two effects, the particles from the evaporating PBHs contribute to the isocurvature perturbations. This enables us to derive new observational constraints on the past abundance of PBHs with $M_\text{PBH} \lesssim 10^{9} \, {\rm g}$ through the CMB constraints on the isocurvature perturbations~\cite{Planck:2018jri}. Our constraints assume primordial non-Gaussianity and hence depends on it, but the PBH isocurvature perturbation itself is a generic phenomenon. PBH bias alone has been used to constrain non-evaporating PBHs through PBH-DM isocurvature modes~\cite{Tada:2015noa, Young:2015kda}, while we apply it to evaporating PBHs, which produces isocurvature modes between all different particle species, photons, baryons, and DM, through evaporation.

%%%%%%%%%%%%%%%%%%%%%%%%%%%%%%%%%%%%%%%%%%%%%%%%%%%%%%%%%%%%%%%%%%%%%%%%%%%%%%%
%%%%%%%%%%%%%%%%%%%%%%%%%%%%%%%%%%%%%%%%%%%%%%%%%%%%%%%%%%%%%%%%%%%%%%%%%%%%%%%
%%%%%%%%%%%%%%%%%%%%%%%%%%%%%%%%%%%%%%%%%%%%%%%%%%%%%%%%%%%%%%%%%%%%%%%%%%%%%%%
\section{Isocurvature perturbations from PBH evaporation}
%%%%%%%%%%%%%%%%%%%%%%%%%%%%%%%%%%%%%%%%%%%%%%%%%%%%%%%%%%%%%%%%%%%%%%%%%%%%%%%
%%%%%%%%%%%%%%%%%%%%%%%%%%%%%%%%%%%%%%%%%%%%%%%%%%%%%%%%%%%%%%%%%%%%%%%%%%%%%%%
%%%%%%%%%%%%%%%%%%%%%%%%%%%%%%%%%%%%%%%%%%%%%%%%%%%%%%%%%%%%%%%%%%%%%%%%%%%%%%%

The current observations on the CMB are consistent with primordial perturbations being predominantly ``adiabatic.'' If all the energy components are from a single origin, they must fluctuate in the same way, ensuring the uniform density hypersurfaces coincide~\cite{Zeldovich:1969sb, Weinberg:2003sw}. The gauge-invariant curvature perturbation on the uniform total density hypersurfaces, $\zeta$, is defined by~\cite{Bardeen:1980kt, Bardeen:1983qw}
\begin{equation}
\label{eq:totalzeta}
\zeta \equiv \varphi - H \frac{\delta\rho}{\dot{\bar\rho}} 
\, ,
\end{equation}
where a bar denotes the background value and $\varphi$ represents the diagonal component of the spatial metric perturbation, given by $\delta g_{ij}\ni 2a^2\varphi\delta_{ij}$. Similarly, the curvature perturbation on uniform $X$-component density hypersurfaces $\zeta_X$ is
\begin{equation}
\zeta_X \equiv \varphi - H \frac{\delta\rho_X}{\dot{\bar\rho}_X}
\, .
\end{equation}
The adiabatic condition means $\zeta_X = \zeta$ for all $X$. More generally, however, $\zeta$ is a weighted sum of the individual $\zeta_X$'s, given by
\begin{equation}
\label{eq:totalzeta-sum}
\zeta = \sum_X \frac{\dot{\bar\rho}_X}{\dot{\bar\rho}} \zeta_X
\, .
\end{equation}
The ``isocurvature'' perturbation quantifies deviations from the adiabatic condition by comparing $\zeta_X$ of any two components
\begin{equation}
    S_{XY} = 3\left(\zeta_X - \zeta_Y\right) 
    = -3 H \left(\frac{\delta \rho_X}{\dot{\bar{\rho}}_X} - \frac{\delta \rho_Y}{\dot{\bar{\rho}}_Y} \right) \label{eq:SXY}\,.
\end{equation}
The constancy of curvature perturbations on superhorizon scales in a universe dominated by a barotropic fluid provides a crucial link between inflationary predictions and observations~\cite{Wands:2000dp, Lyth:2004gb}.
The Planck 2018 results~\cite{Planck:2018jri} have constrained nonadiabatic perturbations to below a few percent at the 95\% confidence level for generic cold DM models. This constraint significantly limits any mechanism for generating isocurvature modes.

To see how evaporation of PBHs generates isocurvature modes, we look at three major components in the early Universe: 
(i) photons ($\gamma$), which include all relativistic particles thermally coupled to photons, (ii) nonrelativistic baryons ($b$), and (iii) cold DM ($d$). We consider the Standard Model neutrinos that remain relativistic and are coupled to photons prior to BBN. Therefore, neutrinos are included in $\gamma$.
The adiabatic condition applies to the curvature perturbations of individual components originating from the same inflationary perturbation, leading to
\begin{equation}
\label{eq:infaadiabatic}
\zeta_{\gamma\text{Inf}} = \zeta_{b\text{Inf}} = \zeta_{d\text{Inf}} = \zeta \, . 
\end{equation}
We focus on large-scale regions accessible by CMB observations. 

On the other hand, PBHs are formed at much smaller horizon scales at earlier epoch. It is the evaporation products of these PBHs that violate the adiabatic condition given in \refeq{infaadiabatic}, leading to the generation of isocurvature modes at large scales. Specifically, Hawking radiation induces curvature perturbations for all three components, given by
\begin{equation}
    \zeta_{\gamma \rm PBH} = \zeta_{b \rm PBH} = \zeta_{d \rm PBH} = \zeta_{\rm PBH}\,. \label{eq:zetaPBHs}
\end{equation}
Upon the completion of PBH evaporation, the final curvature perturbation of component $X$ is a weighted sum of $\zeta_{X\text{Inf}} = \zeta$ and $\zeta_{X\text{PBH}} = \zeta_\text{PBH}$, similar to \refeq{totalzeta-sum}
\begin{equation}
    \zeta_{X, 0} = \frac{\bar{\rho}_{X \rm Inf, 0}}{\bar{\rho}_{X,0}} \zeta + \frac{\bar{\rho}_{X \rm PBH, 0}}{\bar{\rho}_{X,0}} \zeta_{\rm PBH} \, , \label{eq:zetaXA}
\end{equation}
where $\bar{\rho}_{X, 0} = \bar{\rho}_{X \rm Inf, 0} + \bar{\rho}_{X \rm PBH, 0}$. We use the conservation equation $\dot{\bar\rho}_X=-3H(1+w_X)\bar{\rho}_X$ to cancel the time derivative on both sides in the weighting ratio. The subscript 0 denotes the curvature perturbation today, emphasizing that this value determines the final observables. 
Equations~\eqref{eq:SXY} and \eqref{eq:zetaXA} yield the isocurvature perturbation between $X$ and $Y$ as
\begin{equation}
\label{eq:isocurvature}
    S_{XY,0} = 3 \left(\frac{\bar{\rho}_{X \rm PBH, 0}}{\bar{\rho}_{X,0}} - \frac{\bar{\rho}_{Y \rm PBH, 0}}{\bar{\rho}_{Y,0}} \right) (\zeta_{\rm PBH} - \zeta)
    \, .
\end{equation}
Thus, PBH evaporation induces isocurvature perturbations when the following two conditions hold: $\bar\rho_{X\text{PBH},0}/\bar\rho_{X,0} \neq
\bar\rho_{Y\text{PBH},0}/\bar\rho_{Y,0}$ and $\zeta_\text{PBH} \neq \zeta$. The first condition is generically satisfied unless the Hawking radiation branching ratio is finely tuned to match the background abundance, while the second condition holds unless the PBH bias is precisely unity, which is the case when the adiabatic Gaussian perturbation is solely responsible for PBH formation. Therefore, we conclude that the PBH evaporation generically induces isocurvature perturbations between different components of the Universe. Equation~\eqref{eq:isocurvature} is the key equation of our analysis. In the following sections, we examine each condition in greater detail.

%%%%%%%%%%%%%%%%%%%%%%%%%%%%%%%%%%%%%%%%%%%%%%%%%%%%%%%%%%%%%%%%%%%%%%%%%%%%%%%
\subsection{Abundances of particles from PBH evaporation}
%%%%%%%%%%%%%%%%%%%%%%%%%%%%%%%%%%%%%%%%%%%%%%%%%%%%%%%%%%%%%%%%%%%%%%%%%%%%%%%
%%%%%%%%%%%%%%%%%%%%%%%%%%%%%%%%%%%%%%%%%%%%%%%%%%%%%%%%%%%%%%%%%%%%%%%%%%%%%%%%%%%%%%%%%%%%
\begin{figure*}[h!t]
    \centering
    \includegraphics[width = 0.48\linewidth]{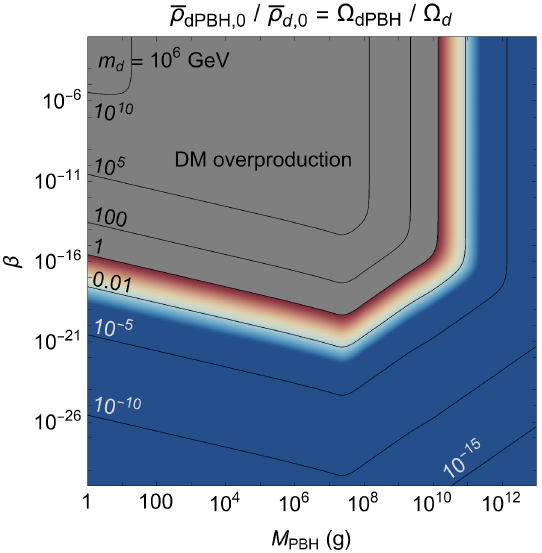}
    \includegraphics[width = 0.48\linewidth]{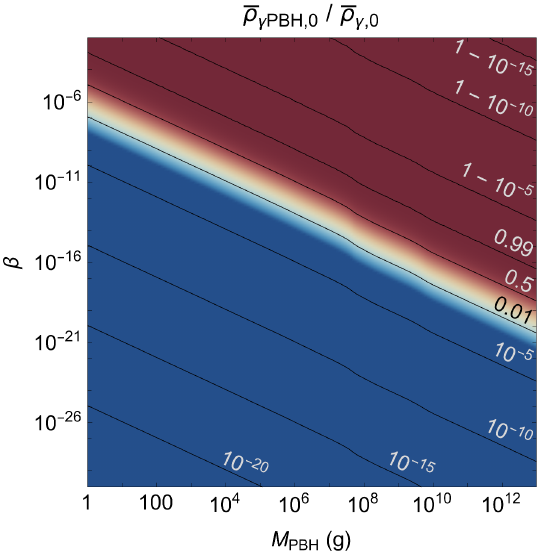}
    \caption{Contour plots of $\bar{\rho}_{d \rm PBH,0}/\bar{\rho}_{d,0}$ for $m_d = 10^6 \, \text{GeV}$ (left) and $\bar{\rho}_{\gamma \rm PBH,0}/\bar{\rho}_{\gamma,0}$ (right) in the parameter space $(M_{\rm PBH}, \beta)$. $\bar{\rho}_{d \rm PBH,0}/\bar{\rho}_{d,0} > 1$ means DM overproduction.}     \label{fig:rhoratio}
\end{figure*}
%%%%%%%%%%%%%%%%%%%%%%%%%%%%%%%%%%%%%%%%%%%%%%%%%%%%%%%%%%%%%%%%%%%%%%%%%%%%%%%%%%%%%%%%%%%%
Evaporating PBHs are characterized by their mass $M_\text{PBH}$ and their initial energy fraction at formation $\beta(M_\text{PBH}) \ll 1$. A PBH of mass $M_{\rm PBH}$ completes its evaporation when the temperature of the Universe reaches
\begin{align}
T_{\rm ev} 
& \approx 
50 \, \text{MeV} \times \left(\frac{M_\text{PBH}}{10^8 \, \text{g}} \right)^{-3/2} 
\nonumber \\
& \quad 
\times 
\left(\frac{g_*(T_{\rm ev})}{10} \right)^{-1/4} \left(\frac{g_H(T_{\rm PBH})}{108} \right)^{1/2}
\,.
\end{align}
$T_{\rm ev}$ also accounts for the reheating of the Universe due to PBH evaporation. Here, $g_*$ represents the entropic relativistic degrees of freedom~\cite{Borsanyi:2016ksw, Husdal:2016haj}, while $g_H$ denotes the Hawking radiation degrees of freedom of the Standard Model particles~\cite{Page:1976df, MacGibbon:1990zk, MacGibbon:1991tj}. The Hawking temperature of a PBH is given by 
\begin{equation}
\label{eq:TPBH}
T_{\rm PBH} = 1.06 \times 10^{5} \, \text{GeV} \times \left(\frac{M_\text{PBH}}{10^8 \, \text{g}}\right)^{-1}
\, . 
\end{equation}
Until evaporation, PBHs behave as pressureless matter, meaning their energy fraction relative to the background radiation increases in proportion to the scale factor. Thus, entropy conservation gives the PBH energy fraction at evaporation as
\begin{equation}
\label{eq:OmegaPBHev}
\Omega_{\rm PBH,ev} 
= 
\left[ 1 + \left( \beta \frac{g_*^{1/3}(T_i) T_i}{g_*^{1/3}(T_{\rm ev}) T_{\rm ev}}\right)^{-1} \right]^{-1}
\, .
\end{equation}
Here, $T_i$ is the temperature at PBH formation given by
\begin{equation}
M_\text{PBH} 
= 
\gamma \frac{4\pi M_{\rm Pl}^2}{H(T_i)} 
\approx 
19 \, \text{g} \times \left(\frac{\gamma}{0.2} \right) \left(\frac{T_i}{10^{15} \, \text{GeV}} \right)^{-2}
\end{equation}
for PBHs produced via standard inflationary mechanism, where $\gamma \simeq 0.2$ represents the fraction of horizon mass that collapses into a PBH~\cite{Carr:1975qj}.

Assuming a universal minimal gravitational coupling to all particles, evaporating PBHs emit blackbody radiation at a temperature $T_{\rm PBH}(M_\text{PBH})$. The branching ratio is directly proportional to the total relativistic degrees of freedom, leading to
\begin{equation}
\label{eq:ratio}
\frac{\bar{\rho}_{X \rm PBH, ev}}{\bar{\rho}_{Y \rm PBH, ev}} 
= 
\frac{\int dT \, T^3 g_{*X}(T)}{\int dT \, T^3 g_{*Y}(T)}
\, .
\end{equation}
This ratio depends on $M_{\rm PBH}$ as the lower bound of the integral is set by $T_{\rm PBH}$ given in Eq.~\eqref{eq:TPBH}. The energy density of each component $\bar{\rho}_{X \rm PBH,ev}$ can then be determined through the normalization $\sum_X \bar{\rho}_{X \rm PBH,ev} = \bar{\rho}_{\rm PBH, ev}$.

For a concrete demonstration, we focus on the case of a monochromatic PBH mass function. This choice simplifies the computation of $\bar{\rho}_{X \rm PBH,ev}$, as PBH evaporation emits particle $X$  when the Hawking temperature satisfies $T_{\rm PBH} \gtrsim m_X/\epsilon_{{\rm em},X_s}$, where $\epsilon_{{\rm em},X_s} = 2.66$, $4.53$, $6.04$, and $9.56$ for spin $0$, $1/2$, $1$, and $2$ particles, respectively~\cite{MacGibbon:1991tj}. We also use the effective Hawking radiation degrees of freedom for particle $X$ with spin $s$, given as $g_{H, X} = g_{s} g_{H,s}$ with $g_{s}$ the number of spin degrees of freedom. The values of $g_{H,s}$ are $g_{H,0} = 1.82$, $g_{H,1/2} = 1.0$, $g_{H,1} = 0.41$, and $g_{H,2} = 0.05$~\cite{Page:1976df, MacGibbon:1990zk}. For $M_{\rm PBH} \sim 10^{13} \, \text{g}$, the sum of all the Standard Model particles gives $g_{H}\sim 100$ and approaches $g_{H}\approx 108$ for lighter $M_{\rm PBH}$.

We introduce two additional simplifications.
First, Hawking evaporation preserves baryon symmetry, meaning that all evaporation products, except for DM, ultimately become photons without generating any net baryon asymmetry. As a result, $\bar{\rho}_{b \rm PBH, 0} = 0$ and we only need to consider photons and DM. Second, we assume a single stable scalar DM that has been out of equilibrium since well before PBH evaporation.

We first consider the DM production from PBH evaporation. If DM particle mass $m_d$ is heavier than $T_{\rm PBH}$, PBHs begin emitting DM particles only when $M_{\rm PBH}$ drops below
\begin{equation}
M_{\rm em} 
= 
1.06 \times 10^8 \, \text{g} \times \epsilon_{{\rm em},d} \left(\frac{10^5 \, \text{GeV}}{m_d} \right)^{-1}
\, . \label{eq:Mem}
\end{equation}
In this case, the majority of emitted particles are nonrelativistic. Conversely, if $m_d$ is lighter than $T_{\rm PBH}$, the emission starts from the beginning and the particles are relativistic over most of the parameter space. At the completion of PBH evaporation, the energy density of DM from PBH is
\begin{align}
\bar{\rho}_{d \rm PBH,ev} 
& = 
\bar{\rho}_{ \rm PBH,ev} 
\nonumber\\
& \quad
    \times \left\{
    \begin{array}{ll}
	\frac{g_{H,d}}{g_{H,d} + g_{H,\rm SM}(T_{\rm PBH,em})} \frac{M_{\rm em}}{M_{\rm PBH}} 
	& (m_d>T_\text{PBH}) \smallskip \\
    \frac{g_{H,d}}{g_{H,d} + g_{H,\rm SM}(T_{\rm PBH})} 
    & (m_d<T_\text{PBH})
    \end{array}
    \right.
    \hspace{-0.4em} .
    \label{eq:rho_dPBH}
\end{align}
This result can be translated into the PBH contribution to the present DM abundance as~\cite{Kim:2023ixo}
\begin{align}
&\Omega_{d \rm PBH,0} 
 \simeq 
10^5 \, \Omega_{\rm PBH,ev} \, g_{H, d} \left(\frac{g_*(T_{\rm ev})}{10} \right)^{1/12} 
\nonumber \\
& \quad
\times 
    \begin{cases}
        1.24 \, \epsilon_{{\rm em}, d} \left(\frac{m_d}{10^5 \, \text{GeV}} \right)^{-1} \left(\frac{M_{\rm PBH}}{10^8 \, \text{g}} \right)^{-5/2}    (m_d>T_\text{PBH})
        \\
        1.10 \, \epsilon_{{\rm em}, d}^{-1} \alpha_d \left(\frac{m_d}{10^5 \, \text{GeV}} \right) \left( \frac{M_{\rm PBH}}{10^8 \, \text{g}}\right)^{-1/2}  (m_d<T_\text{PBH})
    \end{cases}
    \hspace{-1.1em} . 
    \label{eq:Omega_dPBH}
\end{align}
Here, $\alpha_d = 3/2$ for the assumed scalar DM. The present-day DM abundance also accounts for the cosmological redshift of DM particles. 
Finally, we compare \refeq{Omega_dPBH} with the current total DM abundance
\begin{equation}
    \frac{\bar{\rho}_{d \rm PBH,0}}{\bar{\rho}_{d,0}} = \frac{\Omega_{d \rm PBH,0}}{\Omega_{d,0}}
\end{equation}
with $\Omega_{d,0} \simeq 0.26$~\cite{Planck:2018vyg}. Overproduction excludes the corresponding parameter space, while insufficient production implies that the remaining DM must originate from inflationary reheating.

Except for DM, the rest of the PBH evaporation products persist as photons up to the present. Since the number of degrees of freedom for SM particles is much greater than that of DM, Eq.~\eqref{eq:ratio} implies 
$\bar{\rho}_{d \rm PBH,ev} \ll \bar{\rho}_{\rm PBH,ev}$. As a result, the majority of Hawking radiation consists of photons, $\bar{\rho}_{\gamma \rm PBH,ev} \approx \bar{\rho}_{\rm PBH,ev}$. Assuming no significant entropy injection after PBH evaporation, the ratio of photons from Hawking radiation to those from the standard cosmological scenario remains fixed and is given by 
\begin{equation}
\label{eq:gammaratio}
    \frac{\bar{\rho}_{\gamma \rm PBH, 0}}{\bar{\rho}_{\gamma, 0}} \approx \Omega_{\rm PBH,ev} \, . 
\end{equation}
High-energy photons from PBH evaporation thermalize with the surrounding radiation, potentially generating temperature perturbations~\cite{He:2022wwy}. In this work, however, we focus exclusively on the isocurvature perturbations, as described in Eq.~\eqref{eq:isocurvature}.

Figure~\ref{fig:rhoratio} shows the contours of $\bar{\rho}_{d \rm PBH,0}/\bar{\rho}_{d,0}$ and $\bar{\rho}_{\gamma \rm PBH,0}/\bar{\rho}_{\gamma,0}$ in the $(M_{\rm PBH}, \beta)$ parameter space. For a fixed $\beta$, the PBH contribution to DM peaks at $T_{\rm PBH} \sim m_d$, which corresponds to $M_{\rm PBH}\simeq 10^7{\rm g}$ for $m_d=10^6\,{\rm GeV}$. Lighter PBHs with $T_{\rm PBH}>m_d$ evaporate at early times, and their energy density is suppressed due to cosmological redshift. For heavier PBHs, the suppression is primarily due to the factor $M_{\rm em}/M_\text{PBH}$ in \refeq{rho_dPBH}. The vertical upturn of the contour occurs when the PBH density parameter $\Omega_{\rm PBH,ev}$ in Eq.~\eqref{eq:OmegaPBHev} saturates at unity. The gray region in the top-left corner is excluded because DM from Hawking radiation overcloses the Universe. In contrast, the PBH contribution to photons depends monotonically on $\beta$ and $M_{\rm PBH}$, as given in Eqs.~\eqref{eq:OmegaPBHev} and \eqref{eq:gammaratio}. The red region in the top-right marks the parameter space where the majority of photons in the Universe originate from PBH evaporation.

%%%%%%%%%%%%%%%%%%%%%%%%%%%%%%%%%%%%%%%%%%%%%%%%%%%%%%%%%%%%%%%%%%%%%%%%%%%%%%%
\subsection{PBH clustering and bias} %amplitude: $\zeta_{\rm PBH}$}
%%%%%%%%%%%%%%%%%%%%%%%%%%%%%%%%%%%%%%%%%%%%%%%%%%%%%%%%%%%%%%%%%%%%%%%%%%%%%%%

Generally, the curvature perturbation of PBHs differs from the adiabatic one, $\zeta_{\rm PBH} \neq \zeta$, i.e. PBHs are biased. This serves as another key condition for generating isocurvature perturbations through Eq.~\eqref{eq:isocurvature}. 
For example, the PBHs formed from a causal mechanism inside the horizon, such as the phase transitions, scale as $k^2$ on large scales. The local-type primordial non-Gaussianity can also induce the bias of $\zeta_{\rm PBH}$ by systematically altering the small-scale clustering amplitudes $\left<\zeta_s\zeta_s\right>$ in the presence of the long mode $\zeta_\ell$ by $\Delta \left<\zeta_s\zeta_s\right>=\left<\zeta_s\zeta_s\right>|_{\zeta_\ell}$.

The generality of the bias can be understood in the context of effective field theory (EFT). The PBH number density $n_\text{PBH} = n_\text{PBH}(t,\pmb{x}) \propto \beta_\text{PBH}$ is a {\it local} functional dependent on the background cosmology, which could be modulated by any long-wavelength variables. Thus, we may expand $\delta{n}_\text{PBH}/n_\text{PBH}$ in the same manner as the galaxy bias (see e.g.~\cite{Desjacques:2016bnm}) as
\begin{equation}
    \frac{\delta{n}_\text{PBH}}{{n}_\text{PBH}} 
    = 
    \sum_i b_i \mathcal{O}_i + \sum_{i,j} b_{ij} \mathcal{O}_i \mathcal{O}_j + \cdots
    \, ,
\end{equation}
where $\mathcal{O}_i$ is a possible long-wavelength operator and $b_i$ is the corresponding PBH bias, which can be regarded as the Wilsonian coefficient in EFT. Any causal and local operators are allowed, so scalar operators like $\zeta$, and derivative ones such as $\nabla^2\zeta/(aH)^2$ and tidal terms are possible. Thus, nonzero PBH isocurvature perturbations are generic as these allowed operators are not necessarily aligned with $\zeta$. A simple example is that PBH formation depends on a spectator field $\sigma$, which does not contribute to the energy density so that $\zeta$ is not affected. Then, $\beta_\text{PBH} = \beta_\text{PBH}(\sigma)$, so that $\delta{n}_\text{PBH}/n_\text{PBH} = \partial\log\beta_\text{PBH}/\partial\sigma \delta\sigma$, naturally resulting in $\zeta_\text{PBH} \neq \zeta$ even if $\sigma$ is perfectly Gaussian.

Similarly, consider the well-known case of the local-type primordial non-Gaussianity:
\begin{equation}
    \zeta = \zeta_G + \frac{3}{5} f_{\rm NL} (\zeta_G^2 -\langle \zeta_G^2 \rangle) \, , 
\end{equation}
where $\zeta_G$ is Gaussian and $f_{\rm NL}$ parametrizes the lowest-order non-Gaussianity. In terms of EFT, the bias due to $f_{\rm NL}$ corresponds to $\delta{n}_\text{PBH}/n_\text{PBH} \supset b_{\zeta^2}\zeta^2$. 

In the following, for a concrete demonstration, we adopt the linear bias induced by $f_{\rm NL}$ given by~\cite{Tada:2015noa} (but the effect is largely suppressed for a single-field inflation; see Ref.~\cite{Cabass:2016cgp} for example)
\begin{equation}
\label{eq:b}
    b = \frac{6}{5} f_{\rm NL} \left( \frac{\delta_c}{\sigma_s} \right)^2 \, , 
\end{equation}
where $\delta_c \simeq 0.45$ is the density contrast criterion and $\sigma_s$ is the standard deviation of the short-wavelength density perturbation. We once more emphasize that $\zeta_\text{PBH} \neq \zeta$ is very generic and primordial non-Gaussianity needs not be the only possibility. We present the isocurvature bounds based on Eq.~\eqref{eq:b} solely for illustration of our general formula Eq.~\eqref{eq:isocurvature}.
It is straightforward to include higher-order non-Gaussianity, such as $g_{\rm NL}$~\cite{Young:2015kda}.

%%%%%%%%%%%%%%%%%%%%%%%%%%%%%%%%%%%%%%%%%%%%%%%%%%%%%%%%%%%%%%%%%%%%%%%%%%%%%%%
%%%%%%%%%%%%%%%%%%%%%%%%%%%%%%%%%%%%%%%%%%%%%%%%%%%%%%%%%%%%%%%%%%%%%%%%%%%%%%%
%%%%%%%%%%%%%%%%%%%%%%%%%%%%%%%%%%%%%%%%%%%%%%%%%%%%%%%%%%%%%%%%%%%%%%%%%%%%%%%
\section{Isocurvature bounds on evaporating PBHs}
%%%%%%%%%%%%%%%%%%%%%%%%%%%%%%%%%%%%%%%%%%%%%%%%%%%%%%%%%%%%%%%%%%%%%%%%%%%%%%%
%%%%%%%%%%%%%%%%%%%%%%%%%%%%%%%%%%%%%%%%%%%%%%%%%%%%%%%%%%%%%%%%%%%%%%%%%%%%%%%
%%%%%%%%%%%%%%%%%%%%%%%%%%%%%%%%%%%%%%%%%%%%%%%%%%%%%%%%%%%%%%%%%%%%%%%%%%%%%%%
%%%%%%%%%%%%%%%%%%%%%%%%%%%%%%%%%%%%%%%%%%%%%%%%%%%%%%%%%%%%%%%%%%%%%%%%%%%%%%%
\begin{figure}[t]
    \centering
    \includegraphics[width = 0.45\textwidth]{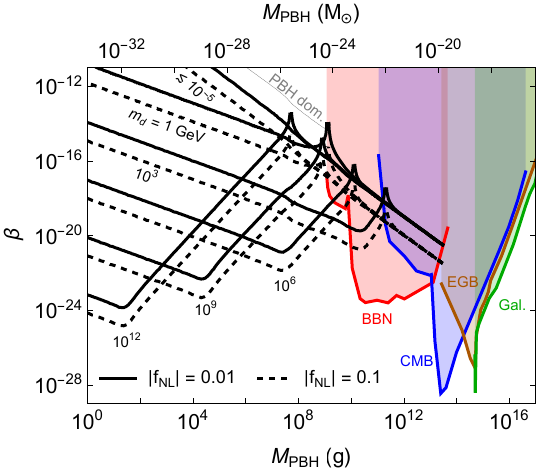}
    \caption{Constraint plot of $\beta(M_\text{PBH})$ from isocurvature perturbations for $|f_{\rm NL}| = 0.01$ (black solid) and $|f_{\rm NL}| = 0.1$ (black dashed), along with bounds from BBN~\cite{Carr:2009jm} (red), CMB~\cite{Carr:2009jm, Acharya:2020jbv, Chluba:2020oip} (blue), extragalactic backgrounds~\cite{Carr:2009jm} (brown), and galactic backgrounds~\cite{Carr:2016hva, Boudaud:2018hqb} (green). 
    Early PBH domination happens above the gray line.}
    \label{fig:betaconstraint}
\end{figure}
%%%%%%%%%%%%%%%%%%%%%%%%%%%%%%%%%%%%%%%%%%%%%%%%%%%%%%%%%%%%%%%%%%%%%%%%%%%%%%%

The Planck 2018 results have placed upper bounds on the isocurvature fraction $\beta_\text{iso}$ in the total power spectrum~\cite{Planck:2013jfk,Planck:2015sxf,Planck:2018jri}
\begin{equation}
\label{eq:betaisofinal}
\beta_{\rm iso} \equiv \frac{\mathcal{P}_S}{\mathcal{P}_\zeta + \mathcal{P}_S}
\, ,
\end{equation}
where $\mathcal{P}_\zeta$ and $\mathcal{P}_S$ denote the power spectra of the adiabatic and isocurvature perturbations, respectively. Here, $S$ represents the total isocurvature perturbation between photons and matter, defined as
\begin{equation}
S \equiv S_{\gamma d,0} + \dfrac{\Omega_{b,0}}{\Omega_{d,0}} S_{\gamma b,0} \,,
\end{equation}
where $\Omega_{b,0}/\Omega_{d,0} \approx 0.1879$~\cite{Planck:2018jri}. 
For $S_{\rm PBH}$ linear in $\zeta$ as given in Eq.~\eqref{eq:b}, $\beta_{\rm iso}$ is independent of both the magnitude of $\zeta$ and wave number $k$, and either fully correlated ($f_{\rm NL} > 0$) or anticorrelated ($f_{\rm NL} < 0$). In Planck 2018, the most stringent constraint is $\beta_{\rm iso} < 0.001$ for all scales for both correlated and anticorrelated cases~\cite{Planck:2018jri}\footnote{For comparison, the most stringent bound without assuming a prior correlation is $\beta_{\rm iso} < 0.025$ at $k = 0.002 \, \text{Mpc}^{-1}$.}.
Figure~\ref{fig:betaconstraint} presents the constraint curves of $\beta(M_{\rm PBH})$ derived from the reported $\beta_{\rm iso}<0.001$ for the PBH mass range $1 \, \text{g} \lesssim M_\text{PBH} \lesssim 3\times 10^{13} \, \text{g}$, with the fiducial value $|f_{\rm NL}|=0.01$ and $0.1$.

For $M_\text{PBH} \lesssim 1 \, \text{g}$, the corresponding inflationary energy scale exceeds the current upper bound set by the constraint on the tensor-to-scalar ratio~\cite{Planck:2018jri}. For $M_\text{PBH} \gtrsim 3\times 10^{13} \, \text{g}$, PBH evaporation occurs after recombination, rendering the induced isocurvature modes unobservable in CMB anisotropies. Between these two limiting cases, the CMB isocurvature constraint imposes a significant upper limit on the PBH energy fraction. Notably, this bound extends to $M_\text{PBH} \lesssim 10^9 \, \text{g}$, a regime previously unconstrained, as such PBHs would have evaporated before BBN.

The constraint strongly depends on the DM particle mass $m_d$. Light DM particles with $m_d\lesssim 10^{-5}\,{\rm GeV}$ remain relativistic too long to constitute a significant fraction of DM today. In this case, similar to $S_{\gamma b,0}$, the isocurvature perturbation $S_{\gamma d,0}$ is determined solely by the excess photons from PBH evaporation, leading to the straight contour of $\beta_{\rm iso}$ following $\Omega_{\rm PBH,ev}$ [see Eq.~\eqref{eq:gammaratio}].
 
In contrast, for $m_d \gtrsim 10^{-5} \, \text{GeV}$, DM from PBHs can constitute a significant fraction of the total DM, leading to nontrivial constraints on $\beta(M_{\rm PBH})$. Here, the most stringent constraint arises when $\bar{\rho}_{\rm dPBH,0}$ is maximized, corresponding to the dips in the left panel of Fig.~\ref{fig:rhoratio} at $T_{\rm PBH}(M_\text{PBH}) \sim m_d$. For larger $M_{\rm PBH}$, the constraint exhibits small peaks that occur when $S_{\gamma d, 0}$ and $S_{\gamma b, 0}$ exactly cancel each other in Eq.~\eqref{eq:betaisofinal}. Only in these narrow parameter regions does the PBH evaporation accidentally restore a perfectly adiabatic matter perturbation, eliminating the isocurvature mode.

In terms of $f_{\rm PBH}$, assuming no evaporation, all of the bounds are placed near $f_{\rm PBH} \sim \mathcal{O} (0.1)$ at their most stringent dips. This is expectable from the bound for nonevaporating PBHs (see Fig.~2 of Ref.~\cite{Tada:2015noa}), which has very mild dependence on $M_{\rm PBH}$ and is placed around $f_{\rm PBH} \sim \mathcal{O} (0.01 \,\text{--} \, 0.1)$. Our isocurvature perturbation has an additional factor arising from the particle composition of evaporation [the first parenthesis of Eq.~\eqref{eq:isocurvature}], which can be $\lesssim 1$ at its maximum, and otherwise follows the contours of Fig.~\ref{fig:rhoratio}. Therefore, the isocurvature bound for evaporating PBHs is a convolution of the bound for nonevaporating PBHs and the particle ratios in Fig.~\ref{fig:rhoratio}.

Above the thin gray line, PBHs induce an early matter-dominated phase~\cite{Inomata:2020lmk, Papanikolaou:2020qtd, Domenech:2020ssp, Domenech:2021wkk, Kim:2024gqp, Holst:2024ubt}. Once PBHs dominate, they become the primary source of the cosmological perturbations, causing $\zeta_{\rm PBH}$ to behave as an adiabatic mode, analogous to the curvaton scenario~\cite{Linde:1996gt, Enqvist:2001zp, Lyth:2001nq, Moroi:2001ct}. This corresponds to $\bar{\rho}_{X \rm PBH,0} / \bar{\rho}_{X,0} \rightarrow 1$ for all $X$'s in Eq.~\eqref{eq:isocurvature}, leading to the suppression of isocurvature modes and thus removing constraints in this regime.

%%%%%%%%%%%%%%%%%%%%%%%%%%%%%%%%%%%%%%%%%%%%%%%%%%%%%%%%%%%%%%%%%%%%%%%%%%%%%%%
%%%%%%%%%%%%%%%%%%%%%%%%%%%%%%%%%%%%%%%%%%%%%%%%%%%%%%%%%%%%%%%%%%%%%%%%%%%%%%%
%%%%%%%%%%%%%%%%%%%%%%%%%%%%%%%%%%%%%%%%%%%%%%%%%%%%%%%%%%%%%%%%%%%%%%%%%%%%%%%
\section{Summary and discussion}
%%%%%%%%%%%%%%%%%%%%%%%%%%%%%%%%%%%%%%%%%%%%%%%%%%%%%%%%%%%%%%%%%%%%%%%%%%%%%%%
%%%%%%%%%%%%%%%%%%%%%%%%%%%%%%%%%%%%%%%%%%%%%%%%%%%%%%%%%%%%%%%%%%%%%%%%%%%%%%%
%%%%%%%%%%%%%%%%%%%%%%%%%%%%%%%%%%%%%%%%%%%%%%%%%%%%%%%%%%%%%%%%%%%%%%%%%%%%%%%

PBHs with $M_{\rm PBH} \lesssim 10^{9} \,{\rm g}$ evaporate before BBN and have thus remained unconstrained until now. In this paper, we demonstrate for the first time that Hawking radiation from these PBHs can induce isocurvature modes observable in the CMB, due to PBH bias and the branching ratio of Hawking radiation. Leveraging this effect, we translate the observed upper bound on isocurvature modes reported by the Planck collaboration into the constraints on the energy fraction of PBHs at their formation. 
As a working example, we consider the primordial non-Gaussianity and present the constraints, which are shown to be very stringent for $f_{\rm NL} \sim \mathcal{O}(10^{-2})$.

Since these light PBHs form at earlier stages of the postinflationary Universe, the constraints derived in this work impose new limits on the amplitude of the small-scale primordial power spectrum, which has never been constrained before. Understanding such small-scale primordial perturbations could provide insights into previously unexplored epochs near the end of inflation or even the onset of reheating after inflation.

Meanwhile, several studies have suggested that a Planckian Hawking temperature or information loss during black hole evaporation may necessitate modifications to the standard Hawking evaporation. At later stages, black holes could leave behind stable remnants instead of complete evaporation~\cite{MacGibbon:1987my, Aharonov:1987tp, Barrow:1992hq, Coleman:1991ku, Banks:1992ba, Banks:1992mi, Banks:1992is, Barrow:1992hq, Casher:1993nr, Carr:1994ar, Alexeyev:2002tg, Chen:2002tu, Barrau:2003xp, Chen:2004ft, Nozari:2005ah, Bai:2019zcd, Lehmann:2019zgt, deFreitasPacheco:2020wdg, Dvali:2020wft, Taylor:2024fvf, Dvali:2024hsb, Davies:2024ysj}. Among these proposals, the memory burden effect~\cite{Dvali:2013eja, Dvali:2020wft, Dvali:2024hsb} has recently gained significant attention. This model predicts a slowdown in black hole evaporation after it has lost half of its mass, potentially leading to stabilization. Many of these scenarios revive the possibility of PBHs as a DM candidate in the previously unconstrained mass range $M_\text{PBH} \lesssim 10^{9} \, \text{g}$.

Several attempts have been made to constrain PBHs in this mass range under the assumption of stable remnants, e.g.~\cite{Carr:1994ar}. In contrast, our isocurvature constraints are independent of such remnant scenarios, as they arise from Hawking radiation, which depends on the nature of DM, the properties of Hawking radiation, and the PBH curvature perturbation $\zeta_{\rm PBH}$. Since all proposed remnant models involve Hawking radiation emission before stabilization, we anticipate that a similar observational constraint could apply to them as well. Modifying our constraints to account for such stable relics is beyond the scope of this work and is left for future study.

\begin{acknowledgments}
\bigskip
\textit{Acknowledgments.}---
We thank Guillem Dom\`enech, Sam Young, Masahide Yamaguchi, Yusuke Yamada, and Tomotaka Kuroda for helpful comments and discussions.
T.H.K. is supported by KIAS Individual Grant No. PG095202 at Korea Institute for Advanced Study.
J.G. is supported in part by the Mid-Career Research Program No. RS-2024-00336507 through the National Research Foundation of Korea Research Grants and by the Ewha Womans University Research Grant of 2024 (No. 1-2024-0651-001-1).
D.J. is supported by NASA ATP Program No. 80NSSC22K0819 and NSF grants (No. AST-2307026, No. AST-2407298) at Pennsylvania State University and by KIAS Individual Grant No. PG088301 at Korea Institute for Advanced Study.
D.W.J. is supported by IBS under the project code IBS-R018-D3.
K.Y.L. is supported by the National Research Foundation of Korea Research Grant No. 2021R1A2C2011003.
J.G. is grateful to the Asia Pacific Center for Theoretical Physics for its hospitality while this work was in progress.

\end{acknowledgments}

\bibliography{apssamp}% Produces the bibliography via BibTeX.

\end{document}